\documentclass[prd,preprintnumbers,nofootinbib]{revtex4}
\usepackage{graphicx}
\usepackage{amsmath}
\usepackage{amsfonts,amsbsy}
\usepackage{amssymb}

\def\gsim{ \,\, \vcenter{\hbox{$\buildrel{\displaystyle >}\over\sim$}}
 \,\,}
\def\lsim{ \,\, \vcenter{\hbox{$\buildrel{\displaystyle <}\over\sim$}}
 \,\,}
\def\be{\begin{equation}}
\def\ee{\end{equation}}
\def\bea{\begin{eqnarray}}
\def\eea{\end{eqnarray}}
\newcommand{\sgn}{\mathop{\mathrm{sgn}}}

\begin{document}

\title{\bf Initial state angular asymmetries in high energy p+A
  collisions: spontaneous breaking of rotational symmetry by a color electric
  field and C-odd fluctuations}

\author{Adrian Dumitru}
\email{adrian.dumitru@baruch.cuny.edu}
\affiliation{Department of Natural Sciences, Baruch College, CUNY,
17 Lexington Avenue, New York, NY 10010, USA}

\author{Andre V. Giannini}
\email{avgiannini@usp.br}
\affiliation{Instituto de F\'isica, Universidade de S\~ao Paulo,
C.P.\ 66318, 05315-970 S\~ao Paulo, SP, Brazil\\
Department of Natural Sciences, Baruch College, CUNY,
17 Lexington Avenue, New York, NY 10010, USA}

\begin{abstract}
We present a simple model for generating initial-state azimuthal
asymmetries in pA collisions from dipole scattering on an anisotropic
dense target. Parity even angular harmonics arise from the C-even real
part of the dipole S-matrix which spontaneously breaks rotational
symmetry in two dimensions due to a condensate for the color electric
field. This leads to an angular correlation with the direction
of $\vec E$.
Parity odd harmonics are generated by the C-odd imaginary part
(odderon) due to coupling to coherent target fluctuations which again
break rotational invariance. We perform a first qualitative extraction
of the amplitude and cutoff of C-odd fluctuations in the dense target.
\end{abstract}

\maketitle

\section{Introduction}

Large azimuthal asymmetries have been observed in p+Pb collisions at
the LHC by the ALICE~\cite{pPb_ALICE},
ATLAS~\cite{pPb_ATLAS,pPb_v1_ATLAS} and CMS~\cite{pPb_CMS}
collaborations. The PHENIX collaboration has also reported similar
asymmetries from d+Au collisions at RHIC~\cite{dAu_RHIC}. These
asymmetries are usually measured via multi-particle angular
correlations (see below) and were found to extend over a long range in
rapidity. By causality, the correlations should therefore originate
from the earliest times of the
collision~\cite{Dumitru:2008wn}. Recently, PHENIX has shown that the
$v_2$ quadrupole moment in the central region of central d+Au
collisions can also be observed in terms of a correlation of particles
with a global ``event plane''~\cite{Adare:2014keg}.

Azimuthal harmonics $v_n=\langle\cos n\phi\rangle$ defined as
$\cos n\phi$ moments of the {\em single-inclusive} distribution
require spontaneous breaking of rotational symmetry in the transverse
plane, defining the event plane. In a classical impact parameter
picture of a binary collision such a preferred direction is provided
by the impact parameter vector. For single-spin asymmetries in
collisions of polarized protons with a heavy ion, the preferred
direction is due to the polarization of the
projectile~\cite{Kovchegov:2012ga}. Here, we explore the scenario that
rotational symmetry is broken by a condensate for the electric field
$\vec E$ in the target~\cite{Kovner:2011pe}, leading to $v_{2n}\ne0$;
and by spontaneous C-odd fluctuations in the
target~\cite{Noronha:2014vva}, leading to $v_{2n+1}\ne0$.

The quadrupole moment $v_2$ in the initial state of high-energy
collisions has been first calculated by Kovchegov and Tuchin long
ago~\cite{Kovchegov:2002nf}. Non-zero $v_2$ emerges due to the
fact that the two-gluon production cross section at relative angle
$\phi_1 -\phi_2\sim 0,\pi$ is enhanced, while it is suppressed for
$\phi_1 -\phi_2\sim \pm \pi/2$. Hence, the two-particle
cumulant~\cite{Borghini:2001vi}
\be \label{eq:v2(2)}
v_2^2\left\{2\right\} = \left< e^{2i(\phi_1-\phi_2)} \right>
\ee
does not vanish. A more recent analysis which also addresses higher
moments $v_n^2\{2\}=\langle e^{in(\phi_1-\phi_2)}\rangle$ of gluons
emitted off the large-$x$ sources has been presented by Gyulassy et
al.~\cite{Gyulassy:2014cfa}.

However, $v_2\{2\}$ fluctuates between events due to
the presence of ``random'' sources of two-particle correlations.
These contributions, if described by a Bessel-Gaussian distribution,
are suppressed by higher-order multi-particle
cumulants~\cite{Borghini:2001vi} such as
\be \label{eq:v2(4)}
v_2^4\left\{4\right\} = 2\left< e^{2i(\phi_1-\phi_3)} \right> \left<
e^{2i(\phi_2-\phi_4)} \right> - \left< e^{2i(\phi_1+\phi_2-\phi_3-\phi_4)} \right>~.
\ee
The fact that $v_2\{2\}\ne0$ or $v_2\{4\}\ne0$ does not imply that
particles are correlated with a global ``event plane''
(i.e.\ spontaneous breaking of rotational symmetry). For a finite
number of particles or sources, multi-particle $v_2$ correlations can
also emerge from fluctuations~\cite{fluc_eps_n}. In this paper,
however, we explore whether the $p_T$ dependence of $v_2(p_T)$
could be consistent with spontaneous breaking of rotational symmetry
due to a condensate for the (color) electric field.

Kovner and Lublinsky have argued the formation of an electric field
condensate at the saturation scale $Q_s$ in large, boosted
nuclei~\cite{Kovner:2011pe}. It is supposed to arise from fluctuations
in the impact parameter dependent small-$x$ evolution in a system with
a correlation length of order $1/Q_s$. If there is a finite
correlation length of order $1/Q_s$ it appears reasonable that over
such distance scales the $\vec E$ field might point in a fixed (but
random) direction. Such an $\vec E$-field condensate spontaneously
breaks rotational symmetry about the beam axis since an incoming
charge is ``deflected'' anisotropically. While the direction of
$\vec E$ in a particular event is of course random, the azimuthal
distribution of the {\em single inclusive} parton cross section with
respect to this preferred direction is not uniform\footnote{In this
  sense, $v_2$ in p+A collisions from a vector target electric field
  is analogous to the ``standard'' $v_2$ from non-central A+A
  collisions where rotational symmetry is spontaneously broken by a
  non-zero impact parameter $\vec b$, which points in a random
  direction.}. The condensate provides a preferred direction for {\em
  any} particle in an event scattering off that particular $\vec E$
field. Here, one of our main goals is to evaluate the $p_T$ dependence
of $v_2$ and $v_4$ for a $\vec r\cdot \vec E$ dipole - electric field
interaction. While we focus on angular harmonics at a fixed rapidity
here, the correlations in fact extend over a long range in rapidity
due to the approximate boost invariance of the target
fields~\cite{Kovner:2010xk}.

We do not account for fluctuation induced $v_2$ (indeed, our $v_{2n}$
vanish in the absence of the condensate) and so our results are more
closely related to $v_2\left\{4\right\}$ rather than
$v_2\left\{2\right\}$. \footnote{We should emphasize again that for a
  finite number of particles, fluctuations can induce two- as well as
  multi-particle correlations in coordinate space~\cite{fluc_eps_n}
  which might then generate correlations in momentum space in the
  final state; this is to be distinguished from the effect studied
  here.} Also, we recall that at leading order in the number of colors
$N_c$, in this approach the $n$-particle distribution factorizes into
$n$ single-particle distributions, each of which exhibits an angular
correlation with the global ``event plane'' defined by the $\vec
E$-field condensate~\cite{Kovner:2010xk}. Connected two-particle
production diagrams~\cite{Dumitru:2008wn,pAridge} appear at relative
order $\sim 1/N_c^2$ and provide corrections to factorization of
angular correlations.

Odd azimuthal harmonics $v_{2n+1}$ may also be induced by C-odd
fluctuations~\cite{Alver:2010gr}. These moments emerge from parity
odd (under $\phi\to\phi+\pi$) contributions to the single-inclusive
distribution, see below. In the dipole picture, they are related to
the C-odd imaginary part of the dipole
S-matrix~\cite{Noronha:2014vva}, just like single-spin asymmetries in
collisions of polarized protons on heavy-ion
targets~\cite{Kovchegov:2012ga,Altinoluk:2014oxa}. Our second main
goal is to provide some first model studies of $v_1$ and $v_3$
(without final state interactions) and to extract an amplitude and
scale for C-odd fluctuations. Just like our calculation of even
harmonics $v_{2n}$, we shall assume that all wave vectors of C-odd
fluctuations point in the same direction and so provide a preferred
``global'' direction for $v_{2n+1}$ which breaks rotational symmetry
spontaneously.

\subsection{Basic setup}
We work in the so-called ``hybrid formalism'' where the proton
projectile is treated as a beam of collinear partons with a large
light-cone momentum $p^-$ which scatter off the field of the
target. To leading order in $p_\perp/p^-$ projectile partons propagate
on eikonal trajectories and the amplitude corresponding to elastic
scattering from momentum $p$ to $q$ is~\cite{BjKS}
\bea
\left< {\rm out}, q| {\rm in}, p\right> &\equiv& \bar{u}(q) \tau(q,p)
u(p) \\
\tau(q,p) &=& 2\pi \delta(p^--q^-) \; \gamma^- \int d^2\vec{x} \left[
  V(\vec{x})-1\right] e^{i (\vec{p}-\vec{q})\cdot \vec{x}}~.
\eea
Here,
\be
V(\vec{x}) = {\cal P} \exp \left[ ig \int dx^- A^+(x^-,\vec{x})\right]
\ee
is a Wilson line along the light cone. Squaring the
amplitude gives the scattering cross section~\cite{Dumitru:2002qt}
\be \label{eq:qA_el}
\frac{d\sigma}{d^2b \,d^2k} =
\frac{dN}{d^2k} =
\frac{1}{(2\pi)^2}
\int d^2r\, e^{-i\vec{k}\cdot \vec{r}}\,
\left(\left< \frac{1}{N_c} {\rm tr}\,\left(
W(\vec{r},\vec{b})
-V(\vec{b}-\vec{r}/2)-V^\dagger(\vec{b}+\vec{r}/2)
\right) \right>+1\right)~.
\ee
Here, $\vec{b}$ denotes the impact parameter of the collision and
$W(\vec{r},\vec{b})$ is a light-like Wilson loop of width given by
$r=|\vec{r}|$. In covariant gauge
$W(\vec{r},\vec{b})=V^\dagger(\vec{b}+\vec{r}/2) \,
V(\vec{b}-\vec{r}/2)$, commonly referred to as the dipole
unintegrated gluon distribution~\cite{MuellerDipole}. The size of the
dipole is given by the shift of the transverse coordinate of the
eikonal quark line from the amplitude to the complex conjugate
amplitude, respectively.  The expression in parenthesis corresponds to
the S-matrix for a dipole of size and orientation given by $\vec r$.

Averaging over a C-even ensemble of target fields in
eq.\ (\ref{eq:qA_el}) provides the real part $D(\vec r)$ of the dipole
S-matrix which is even under $\vec r\to-\vec r$, see below. On the
other hand, C-odd fluctuations provide an expectation value for the
``odderon'' Im$\,S=O(\vec r)$ which is odd under $\vec r\to-\vec r$.
Eq.\ (\ref{eq:qA_el}) can be turned into a physical $pA\to h+X$ single
inclusive cross section for production of a hadron of type $h$ via a
convolution with a proton-parton distribution and a corresponding
$q\to h$ fragmentation
function~\cite{Dumitru:2005gt,Altinoluk:2011qy,Chirilli:2011km}. The
present paper does not aim at a quantitative comparison to the data
and so we presently drop these convolutions with the parton
distribution and fragmentation functions. Our more modest goal is to
obtain some basic understanding of the behavior of the azimuthal
harmonics $v_n$ from eq.~(\ref{eq:qA_el}).

We should stress that while eq.~(\ref{eq:qA_el}) does resum coherent
multiple scattering of a projectile parton with the target field, it
intrinsically assumes that the projectile is dilute and that
non-linear (high density) effects in the proton can be neglected. This
may not be a good representation of high multiplicity p+A collisions
at LHC energies (except towards the fragmentation region of the
proton~\cite{Dumitru:2002qt}). Small-$x$ evolution effects of the
proton can in principle be treated, see e.g.\ refs.~\cite{pAridge}
specifically for studies of the ``ridge'' resp.\ of $v_2$ in the
fluctuation dominated regime. A related issue is that non-trivial
field configurations such as semi-hard ``strings'' can form in
collisions of dense sheets of color charge, and their decay may also
induce angular correlations~\cite{Andres:2014bia}.  However, given the
relative simplicity with which we can incorporate an electric field
condensate in eq.~(\ref{eq:qA_el}) we find it interesting to explore
the angular distribution it predicts, and in particular to show how
parity odd angular moments of the single-inclusive distribution are
generated by fluctuations of the target field.

\section{Transverse momentum distribution of scattered quarks}

The transverse momentum distribution of quarks scattered off the target
can now be written as\footnote{Eq.~(\ref{eq:Def_dN0}) includes the
  ``no scattering'' contribution for transverse momentum exchange
  $k=0$. It plays no role in our subsequent analysis since we are
  interested in finite $k$ only.}
\bea
(2\pi)^2\,
\frac{dN}{k dk \, d\phi_k}
&=& \int d^2r\, e^{-i \vec k \cdot \vec r} \; S(\vec r) \label{eq:Def_dN0}\\
&=& \int dr\; r\, d\phi_r\, e^{-i kr\cos(
\phi_k-\phi_r)} \, S(r,\phi_r) \label{eq:Def_dN1}~,
\eea
with $S(\vec{r})$ the S-matrix of a dipole of size and orientation
given by $\vec r$.  The transverse momentum distribution is a real
function, hence
\be
S(r,\phi_r) = S^*(r,\phi_r+\pi)~.
\ee
In particular, the real and imaginary parts of $S(\vec{r})$ satisfy
\bea
S(r,\phi_r) &=& D(r,\phi_r) + i O(r,\phi_r)~,  \label{eq:S_D_O}\\
D(r,\phi_r) &=& D(r,\phi_r+\pi)~, \label{eq:D_Peven} \\
O(r,\phi_r) &=& -O(r,\phi_r+\pi)~.
\eea
Thus, the real part $D(\vec{r})$ is even under $\phi_r\to\phi_r+\pi$
(i.e.\ $\vec{r}\to-\vec{r}$) while $O(\vec{r})$ is odd.

The S-matrix for a dipole in the adjoint representation on the other
hand is real. Written in terms of the real and imaginary parts of the
fundamental representation from eq.~(\ref{eq:S_D_O}) it is
\be
S_{\rm adj}(\vec r) = \frac{N_c^2[D^2(\vec r)+O^2(\vec r)]-1}{N_c^2-1}
\rightarrow D^2(\vec r)+O^2(\vec r)~,
\ee
where the right most expression applies in the large-$N_c$
limit. Clearly $S_{\rm adj}(\vec r) = S_{\rm adj}(-\vec r)$ and thus
can only generate non-zero $v_{2n}$; this has been pointed out before
in ref.~\cite{Kovner:2010xk}.

\section{Azimuthal harmonics $v_n$}

We can define various asymmetry moments $v_n$ through
\be \label{eq:Def_vn}
v_n(k_T) = \left< \cos n\phi_k \right> = \frac{1}{\cal N}
\int \frac{d\phi_k}{2\pi} \cos (n\phi_k)\, \frac{dN}{dy\, k_T dk_T \, d\phi_k}~,
\ee
with
\be \label{eq:Norm_kt}
{\cal N} = \int \frac{d\phi_k}{2\pi} \; \frac{dN}{k_T dk_T \,
  d\phi_k} = \frac{1}{\pi} \frac{dN}{dk_T^2}~.
\ee

Even (odd) moments have positive (negative) parity:
\bea
\left< \cos 2n\phi_k \right> &=& + \left< \cos 2n(\phi_k+\pi) \right> ~,\\
\left< \cos (2n+1)\phi_k \right> &=& - \left< \cos (2n+1)(\phi_k+\pi) \right> ~.
\eea

If the scattering amplitude $S(r,\phi_r)$ is independent of the
orientation of the dipole then all $v_n=0$. An angular dependence of
its real part $D(r,\phi_r)$ gives rise to non-zero parity even moments
$v_{2n}$; an angular dependence of $O(r,\phi_r)$ produces odd moments
$v_{2n+1}$.

In the following sections we employ simple schematic models for the
dipole S-matrix to work out qualitative features of the azimuthal
moments $v_n(k_T)$.

\section{Models for the real part of the dipole scattering amplitude}

\subsection{Quasi-classical dipole model} \label{sec:MVmodel}

In the classical McLerran-Venugopalan model of Gaussian color charge
fluctuations the real part of the dipole scattering amplitude is given
by~\cite{Kovchegov:1998bi}
\be \label{eq:avgD}
D(r) = e^{- \frac{1}{4} r^2 Q_s^2 \log \frac{1}{\Lambda r}}~.
\ee
To arrive at this expression one has averaged over all configurations
of the target and so $D(\vec r)$ is rotationally symmetric. With this
rotationally symmetric dipole the transverse momentum distribution at
$k_T\gg Q_s$ becomes
\be \label{eq:unpolMV_highk}
\frac{dN}{k_T dk_T \, d\phi_k} = \frac{1}{2\pi} \frac{Q_s^2}{k_T^4}
+ \cdots
\ee
Following Kovner and Lublinsky~\cite{Kovner:2011pe} we instead
consider an average over the color charge configurations of the target
at {\em fixed} relative angle of dipole and target field. That is,
local rotational symmetry in the transverse plane is spontaneously
broken in a particular event by the direction of $E^i=F^{+i}$ within
the domain. The presence of such a condensate has been argued to
provide a possible explanation for the large $v_2\{4\}$ observed in
high multiplicity p+Pb collisions at the LHC, see section~3 in
ref.~\cite{Dumitru:2013tja}.

We perform a global rotation of the event such that $\vec E = (E,0)$
points in the $\vec x$-direction\footnote{This is according to the
  convention that the resulting transverse momentum distribution is
  symmetric under $\phi_k\to-\phi_k$, and that $v_2\equiv\langle \cos
  2\phi_k\rangle$ is maximized.}.  We then have $(\vec r\cdot \vec E)^2
= E^2 r^2 \cos^2\phi_r$. Hence
\bea
D(\vec r) &=&
e^{- \frac{1}{4} r^2 Q_s^2 \, (1-{\cal A}+2{\cal A}
\cos^2 \phi_r) \, \log \frac{1}{\Lambda r}}~, \label{eq:D_E}
\eea
where $\cal A$ determines the degree of polarization of the target
field, with ${\cal A}=1$ corresponding to perfect polarization while
${\cal A}=0$ leads back to the unpolarized target from
eq.~(\ref{eq:avgD}). $D(\vec r)$ clearly satisfies~(\ref{eq:D_Peven})
from above, i.e.\ it has even parity under $\vec r\to -\vec r$, and so
all $v_{2n+1}=0$. Note that if
\be
\frac{g^2}{N_c}\sum\limits_{a,b} \left< \vec r\cdot{\vec E}^{a}\;
\vec r\cdot{\vec E}^{b} \right>
\sim r^2Q_s^2\left(1-{\cal A}+2{\cal
  A}\cos^2 \phi_r\right)
\ee
with the same angle $\phi_r$ for all color channels
then ${\cal A}={\cal O}(1)$. On the other hand, if the direction of
${\vec E}^a$ in different color channels fluctuates independently then
${\cal A}$ is suppressed by the appropriate power of $N_c$.

We emphasize that in~(\ref{eq:D_E}) we have not
explicitly averaged over multiple domains of $\vec E$ as described in
ref.~\cite{Kovner:2011pe} but in essence assume scattering off a
single $\vec E$ domain with effective polarization ${\cal
  A}$. We present a simple domain model in
section~\ref{sec:Domain_Model}.

The transverse momentum distribution of quarks scattered to $k_T\gg
Q_s$ now is
\be \label{eq:dNdkt_phi_high_kT}
\frac{dN}{k_T dk_T \, d\phi_k} =
\frac{1}{2\pi}
\frac{Q_s^2}{k_T^4} \, \left[1 - 2{\cal A}(k_T)+4{\cal A}(k_T)
\cos^2 \phi_k\right]~.
\ee
Because of the absence of non-linear effects for the classical
dipole~(\ref{eq:D_E}) at high $k_T$, the angular average ($\langle
\cos^2\phi_k\rangle =1/2$) leads back to the transverse momentum
distribution for an unpolarized target from
eq.~(\ref{eq:unpolMV_highk}).

The angular distribution~(\ref{eq:dNdkt_phi_high_kT}) leads to
\be
v_2(k_T) \equiv \langle \cos 2\phi_k\rangle =
{\cal A}(k_T)~,  \label{eq:v2_highk_MV}
\ee
so that $v_2(k_T)$ is related to the target field polarization at the
scale $k_T$, with all other $v_{n\ne 2}=0$. One may expect that beyond
the saturation scale ${\cal A}(k_T)$ could decrease with increasing
transverse momentum since the short wavelength modes of $\vec E$
should have random orientations. In our numerical estimates below for
simplicity we shall assume ${\cal A}$=const.

We can also obtain the Fourier transform of~(\ref{eq:D_E}) at $k_T\ll
Q_s$. Here, the maximal dipole size $r$ is not set to $\sim 1/k_T$ by
the Fourier phase but to $\sim 1/Q_s$ by saturation. The transverse momentum
distribution becomes
\be
\label{eq:dNdkt_phi_low_kT}
\frac{dN}{k_T dk_T \, d\phi_k} =
\frac{1}{\pi Q_s^2\log Q_s/\Lambda}\frac{1}{\sqrt{1-{\cal A}^2}}
\exp\left[ - \frac{k_T^2}{Q_s^2 \log Q_s/\Lambda}
\left(\frac{\cos^2\phi_k}{1+{\cal A}} + \frac{\sin^2\phi_k}{1-{\cal
    A}}
\right) \right]~.
\ee
To obtain the normalization factor we average over $\phi_k$:
\be
\label{eq:dNdkt_phi_low_kT_norm}
\frac{1}{\pi} \frac{dN}{dk_T^2} =
\frac{1}{\pi Q_s^2\log Q_s/\Lambda} \frac{1}{\sqrt{1-{\cal A}^2}}\,
I_0\left( \frac{{\cal A}}{1-{\cal A}^2}\frac{k_T^2}{Q_s^2 \log
    Q_s/\Lambda}\right)\,
\exp\left[ - \frac{k_T^2}{Q_s^2 \log Q_s/\Lambda} \frac{1}{1-{\cal
      A}^2} \right]~.
\ee
Hence, the normalized angular distribution is given by
\be \label{eq:AngDis_low_kT}
\frac{1}{I_0\left(\frac{{\cal A}}{1-{\cal A}^2}\frac{k_T^2}{Q_s^2 \log
    Q_s/\Lambda}\right)}\,
\exp \left[
\frac{k_T^2}{Q_s^2 \log Q_s/\Lambda} \frac{{\cal A}}{1-{\cal A}^2}
\left( \cos^2\phi_k- \sin^2\phi_k\right)\right]~.
\ee
Here, the polarization amplitude ${\cal A}={\cal A}(Q_s)$ is measured
at the scale $Q_s$ and is therefore independent of
$k_T$. Eq.~(\ref{eq:AngDis_low_kT}) leads to
\bea
v_2(k_T) &=& \frac{I_1\left( \frac{{\cal A}}{1-{\cal A}^2}\frac{k_T^2}{Q_s^2 \log
    Q_s/\Lambda}  \right)}
{I_0\left(\frac{{\cal A}}{1-{\cal A}^2}\frac{k_T^2}{Q_s^2 \log
    Q_s/\Lambda}\right)} \simeq
\frac{{\cal A}}{2} \; \frac{k_T^2}{Q_s^2 \log
    Q_s/\Lambda} \label{eq:v2_lowk_gamma1}\\
v_4(k_T) &=& \frac{I_2\left( \frac{{\cal A}}{1-{\cal A}^2}\frac{k_T^2}{Q_s^2 \log
    Q_s/\Lambda}  \right)}
{I_0\left(\frac{{\cal A}}{1-{\cal A}^2}\frac{k_T^2}{Q_s^2 \log
    Q_s/\Lambda}\right)} \simeq
\frac{{\cal A}^2}{8} \; \left(\frac{k_T^2}{Q_s^2 \log
    Q_s/\Lambda}\right)^2~.  \label{eq:v4_lowk_gamma1}
\eea
The expressions on the right hand side apply when the polarization
amplitude of the target field ${\cal A}\ll 1$. Thus, $v_4(k_T)$ is not
zero in the saturation regime though smaller than $v_2(k_T)$ by one
additional power of ${\cal A}$ and of $k_T^2/Q_s^2$.

\subsection{Dipole with anomalous dimension}
The dipole aquires an anomalous dimension $\gamma<1$ upon
resummation of quantum fluctuations in the small-$x$ regime:
\bea
D(\vec r) &=&
e^{- \left(\frac{1}{4} r^2 Q_s^2 \, (1-{\cal A}+2{\cal A}
\cos^2 \phi_r) \right)^\gamma~}. \label{eq:D_E_gamma}
\eea
(Logarithms of $1/r$ do not affect the transverse momentum
distribution qualitatively when $\gamma<1$.)

In this subsection we discuss the corresponding angular distribution
and its moments $v_n$. We shall restrict to $\gamma=1/2$ in order to
derive simple analytical expressions. This should suffice to
illustrate the qualitative effect of the anomalous dimension on $v_n$.

In the ${\cal A}=0$ isotropic
limit the transverse momentum distribution takes the form
\be
\frac{1}{\pi} \frac{dN}{dk_T^2} = \frac{1}{4\pi}\,
\frac{Q_s}{\left( k_T^2 + Q_s^2\right)^{3/2}}~.
\ee
In the presence of a polarized condensate, i.e.\ for ${\cal A}>0$, we
consider the limits $k_T\gg Q_s$ and $k_T\ll Q_s$ separately. At high
transverse momentum we can expand~(\ref{eq:D_E_gamma}) in powers of
$r$. The leading contribution $\sim r$, for ${\cal A}\ll1$, gives
\be
\frac{dN}{d^2k_T} = \left(1-\frac{3{\cal A}}{2} +
3{\cal A}\cos^2\phi_k\right)
\frac{1}{4\pi}\frac{Q_s}{k_T^3}
~~~~~~~~~(k_T\gg Q_s,~{\cal A}\ll1)~.
\ee
From this expression we find that
\bea
v_2(k_T) &=& \frac{3}{4} {\cal A}(k_T)~, \label{eq:v2_highk_gamma1/2}\\
v_{2n\ge4} &=& 0~.   \label{eq:v4_highk_gamma1/2}
\eea
Hence, for $k_T\gg Q_s$ we find that the anomalous dimension $\gamma$
does not affect $v_2(k_T)$ and $v_4(k_T)$ qualitatively.

For $k_T\ll Q_s$ we start by writing the transverse momentum
distribution in the form
\be
\frac{dN}{d^2k_T} = \frac{1}{\pi}\, \int\limits_{-\pi}^\pi d\phi_r
\frac{1}{\left(
Q_s \sqrt{1+{\cal A}\cos 2\phi_r} + 2 i k_T \cos \left(\phi_r-\phi_k\right)
\right)^2}~.
\ee
Expanding in powers of $k_T$ we find that
\be
\frac{dN}{d^2k_T} = \frac{1}{Q_s^2} \left(
\frac{2}{\sqrt{1-{\cal A}^2}} - 12
\frac{k_T^2}{Q_s^2} \, \frac{1-{\cal A}\cos 2\phi_k}{(1-{\cal
    A}^2)^{3/2}}\right)~.
\ee
Averaging this expression over $\phi_k$ gives the normalization factor
\be
{\cal N} = \frac{1}{Q_s^2} \left(
\frac{2}{\sqrt{1-{\cal A}^2}} - 12\frac{k_T^2}{Q_s^2} \,
\frac{1}{(1-{\cal A}^2)^{3/2}}\right)~.
\ee
Thus,
\bea \label{eq:v2_lowk_gamma1/2}
v_2(k_T) &=& \frac{6}{\cal N}\, \frac{k_T^2}{Q_s^4}
\frac{{\cal A}}{(1-{\cal A}^2)^{3/2}} =
\frac{3 {\cal A} k_T^2} {(1-{\cal A}^2)Q_s^2 - 6 k_T^2}~, \\
v_4(k_T) &=& 0~.
\eea
Note that this expression is valid only as long as ${\cal N}>0$,
i.e.\ for $k_T^2 \ll (1-{\cal A}^2)Q_s^2/6$. Hence, for $\gamma=1/2$
we find that $v_4$ (and all higher harmonics) vanish at low $k_T$, in
contrast to the classical model where $v_4\sim {\cal A}^2\, k_T^4$,
eq.~(\ref{eq:v4_lowk_gamma1}).

Table~\ref{tab:v2v4} summarizes our results from
eqs.~(\ref{eq:v2_highk_MV}, \ref{eq:v2_lowk_gamma1},
\ref{eq:v4_lowk_gamma1}, \ref{eq:v2_highk_gamma1/2},
\ref{eq:v4_highk_gamma1/2}, \ref{eq:v2_lowk_gamma1/2}).
\begin {table}[htb]
\begin{center}
\begin{tabular}{ c|c|c|c }
 \hline
             & $v_2$            & $v_4$ & $\gamma$\\
\hline
 $k_T\gg Q_s$ & ${\cal A}(k_T)$      & 0 & $\gamma\to1$\\
             & ${\cal A}(k_T)$       & 0 & $\gamma = 1/2$\\
\hline
 $k_T\ll Q_s$ & ${\cal A}\,k_T^2$ & ${\cal A}^2\,k_T^4$ & $\gamma\to1$ \\
              & ${\cal A}\,k_T^2$ & 0                  & $\gamma=1/2$ \\
 \hline
\end{tabular}
\caption{Summary of the leading dependence on $k_T$ and condensate amplitude
  ${\cal A}$ of $v_2$ and $v_4$.}  \label{tab:v2v4}
\end{center}
\end{table}

\section{Models for the imaginary part of the dipole scattering amplitude}

In this section we consider the angular distribution due to the
imaginary part of the dipole forward scattering amplitude. We restrict
to the semi-classical approximation and take~\cite{Kovchegov:2012ga}
\be \label{eq:iO_gradD}
iO(\vec r) \sim i \vec r \cdot \vec\nabla_b D(\vec r, \vec b)~.
\ee
The proportionality constant is given in ref.~\cite{Kovchegov:2012ga}
and shall be restored later. Here $\vec b$ denotes the
transverse coordinate of the dipole center of mass (relative to the
center of the nucleus): $\vec b = (\vec x +\vec y)/2$ with $\vec x$
and $\vec y$ the endpoints of the dipole.  $D(\vec r, \vec b)$ depends
on $\vec b$ through $Q_s^2(\vec b)$. Throughout the manuscript we work
in the approximation where the nucleus is infinite and homogeneous
{\em on average} so that the average $Q_s^2(\vec b) = Q_s^2$ is
constant. However, the projectile dipole can still couple to short
wavelength fluctuations in the target and provide a non-vanishing
odderon contribution $O(\vec r)$. To account for coupling to
fluctuations at the scale $r$ we generalize~(\ref{eq:iO_gradD}) to
\be \label{eq:iO_gradient_string}
iO(\vec r) \sim i \int\limits_{\vec y}^{\vec x} d\vec s
\cdot \vec\nabla D(\vec r, \vec s)~.
\ee
This resembles the situation encountered in ref~\cite{Noronha:2014vva}
where the dipole S-matrix was computed from the AdS/CFT
correspondence: the Nambu-Goto action involves an integration of the
(fluctuating) target density along the string connecting the quark and
anti-quark. In case that the fluctuations occur on scales larger than
$\sim r$, eq.~(\ref{eq:iO_gradient_string}) returns
to~(\ref{eq:iO_gradD}) as the average gradient along the dipole can
then be taken at the midpoint $\vec b$.

We begin by assuming random polarization of electric fields in the
target, so ${\cal A}=0$ in the notation of the previous section; the
effect of the $\vec E$ field condensate will be discussed
below. In the classical field limit then
\be
iO(\vec r) \sim - \frac{i}{4} r^2 \log \frac{1}{\Lambda r} \,
e^{- \frac{1}{4} r^2 Q_s^2 \log \frac{1}{\Lambda r}} \int\limits_{\vec y}^{\vec x} d\vec s
\cdot \vec\nabla Q_s^2(\vec s)~.
\ee
If the fluctuation in the target is dominated by a single mode we may
write
\bea
\frac{Q_s^2(\vec s)}{Q_s^2} &=& 1 + \int \frac{d^2q}{(2\pi)^2} \, \delta
f(\vec q) \, e^{i \vec q\cdot \vec s} \\
\delta f(\vec q) &=& \frac{(2\pi)^2}{2} {\cal B}(q_0) \left[
(1+i)\delta(\vec q - \vec q_0) + (1-i)\delta(\vec q + \vec q_0)
\right]~.
\eea
Here, $\vec q_0=(q_0,0)$ determines the scale of the fluctuation and
the direction for spontaneous breaking of rotational symmetry. ${\cal
  B}(q_0)$ is the amplitude of the fluctuation. Note that the
fluctuation satisfies $\delta f(\vec q)=\delta f^*(-\vec q)$ so that
$Q_s^2(\vec s)$ is real. We should stress that the direction of the
wave vector $\vec q$ of the fluctuation would, in principle, be
arbitrary and that it need not coincide with the direction of the
condensate which generates the even harmonics. However, here we shall
not be concerned with correlations of the ``event planes'' of $v_n$
and so we fix $\vec q=(q_0,0)$ to point in the same direction as the
$\vec E$ field condensate discussed in the previous section.

With this {\it ansatz},
\bea
 \int\limits_{\vec y}^{\vec x} d\vec s\cdot \vec\nabla Q_s^2(\vec s)
&=& - Q_s^2\,{\cal B}(q_0) \int\limits_{\vec y}^{\vec x} d\vec s\cdot \vec q_0 \,
\left[ \sin\left(\vec q_0\cdot\vec s\right)
+ \cos\left(\vec q_0\cdot\vec s\right)
\right]~, \\
&=&
- 2 Q_s^2\,{\cal B}(q_0) \, \sin\frac{\vec q_0\cdot\vec r}{2}~,
\eea
where we parametrized the line from $\vec y$ to $\vec x$ as $\vec
s(\sigma) = \vec b + \vec r\, \sigma$ with
$\sigma\in[-\frac{1}{2},\frac{1}{2}]$, and then set $\vec b = 0$.

Hence, we obtain
\bea  \label{eq:iO_all_r}
iO(\vec r) &\sim& \frac{i}{2} \, r^2 \, Q_s^2 \, {\cal B}(q_0)
\sin\left( \frac{1}{2}
rq_0 \cos\phi_r\right) \log \frac{1}{\Lambda r}\, e^{- \frac{1}{4} r^2 Q_s^2 \log
  \frac{1}{\Lambda r}} ~.
\eea

To obtain some qualitative insight consider the limit of high
transverse momentum resp.\ $r \ll 1/Q_s$. In this regime we can
drop the exponential in~(\ref{eq:iO_all_r}):
\bea  \label{eq:iO_r<<1/Qs}
iO(\vec r) &\sim& \frac{i}{2} \, r^2 \, Q_s^2 \, {\cal B}(q_0)
\sin\left( \frac{1}{2}
rq_0 \cos\phi_r\right)  \log \frac{1}{\Lambda r}~~~~~~~~~~~~~~~
(rQ_s \ll 1)~.
\eea
A very long wavelength fluctuation (as compared to the size of the
dipole) corresponds to $r q_0\ll1$ and the leading approximation for
the sine leads to
\be
\frac{dN}{d^2k_T} \sim {\cal B}(q_0)\, \frac{q_0}{k_T}
\frac{Q_s^2}{k_T^4} \cos(\phi_k)~.
\ee
In fact, the limit $r q_0\ll1$ corresponds to evaluating the gradient
of $D(\vec{x},\vec{y})$ at the midpoint $\vec b$ as in
eq.~(\ref{eq:iO_gradD}). This angular distribution evidently
corresponds to $v_1(k_T) \ne 0$ but $v_3(k_T)=0$.

Fluctuations at the scale $r q_0\gsim1$ do lead to higher harmonics,
however. To show that we integrate eq.~(\ref{eq:iO_all_r}) over a
scale invariant spectrum of fluctuations:
\be
{\cal B}(q_0) \rightarrow \frac{1}{2} {\cal B}
 \int\frac{dq_0^2}{q_0^2}  ~,
\ee
acting on the sine function in that expression. Note that in doing so,
we are still restricting to the somewhat extreme case where all wave
vectors $\vec q_0$ are pointing in the same direction; some angular
smearing of the fluctuation vectors about the $\vec x$-direction could
be incorporated in the future. This turns eq.~(\ref{eq:iO_all_r}) into
\be \label{eq:iO_fluc_1/q02}
iO(\vec r) \sim \frac{i \pi}{4} \, r^2 \, Q_s^2 \,{\cal B} \,
\sgn \left(\cos\phi_r\right)
\log \frac{1}{\Lambda r}\, e^{- \frac{1}{4} r^2 Q_s^2 \log
  \frac{1}{\Lambda r}} ~.
\ee

We can generalize this equation by adding an exponential fluctuation
cutoff beyond a scale $Q_c$:
\be
{\cal B}(q_0) \rightarrow \frac{1}{2} {\cal B}
 \int\frac{dq_0^2}{q_0^2}  \, e^{- q_0/Q_c}~,
\ee
which again is acting on the sine function in
eq.~(\ref{eq:iO_all_r}). Then the C-odd part of the S-matrix is
\be \label{eq:iO_fluc_1/q02_cutoff}
iO(\vec r) \sim \frac{i}{2} \, r^2 \, Q_s^2 \,{\cal B} \,
\arctan \left(\frac{1}{2}rQ_c\cos\phi_r\right)
\log \frac{1}{\Lambda r}\, e^{- \frac{1}{4} r^2 Q_s^2 \log
  \frac{1}{\Lambda r}} ~.
\ee
Eq.~(\ref{eq:iO_fluc_1/q02}) is reproduced when $Q_c\to\infty$. The
expansion for small dipoles, $r^{-1}\gg Q_c,\, Q_s$ is
\be
iO(\vec r) \sim i \, r^3\, Q_s^2 \, Q_c\, {\cal B} \cos \phi_r \left[
1 - \frac{r^2}{4} \left( \frac{Q_c^2\cos^2 \phi_r}{3} + Q_s^2
  \right)
\right]~,
\ee
up to logarithms of $1/r$. Hence, $v_1$ is generated already at order
$r^3$ while $v_3$ only appears at order $r^5$ and would therefore drop
by an additional $1/k_T^2$ at high transverse momentum.

Finally, we restore the proper prefactor of $iO(\vec r)$ as given in
ref.~\cite{Kovchegov:2012ga}:
\be \label{eq:iO_fluc_1/q02_norm}
iO(\vec r) = i \, \frac{\alpha_s \, (N_c^2-4)}{2^6}
\, {\cal B} \,
\arctan \left(\frac{1}{2}rQ_c\cos\phi_r\right)
\, \rho^2\, e^{-\rho^2} ~,
\ee
with
\be \label{eq:rho2_A=0}
\rho^2 = \frac{1}{4}\, r^2 \, Q_s^2 \, \log \frac{1}{\Lambda r}~.
\ee
The effect of the $\vec E$ field condensate would be to modify this
expression to
\be  \label{eq:rho2_A>0}
\rho^2_{\cal A} = \rho^2\, \left( 1 - {\cal A} +
2{\cal A} \cos^2 \phi_r \right) ~.
\ee
Eq.~(\ref{eq:rho2_A>0}) assumes that the fluctuations which generate
odd harmonics are perfectly aligned with the $\vec E$ field condensate
which generates even $v_{2n}$. For simplicity, we shall assume that this
is not the case and instead take the event planes of $v_{2n}$ and
$v_{2n+1}$ to have random relative orientations. Thus, below we shall
use eq.~(\ref{eq:rho2_A=0}).

\section{Numerical Results}

In this section we present numerical results for the azimuthal
harmonics $v_n$. We have discretized $\vec r$-space on a
two-dimensional cartesian lattice, performed a Fourier transform to
$2d$ discrete $\vec k$-space where we then evaluated
\be \label{eq:vn(k)}
v_n(k) = e^{i\pi\delta_{n1}} \left(2\frac{dN}{dk^2}\right)^{-1}
\int d\phi_k
  \frac{dN}{d^2k}\cos n \phi_k ~.
\ee
Even if $dN/d^2k$ is chosen to be symmetric under $\phi_k\to-\phi_k$
there is, in general, an undetermined phase $\exp (i\pi j)$ with integer
$j$ which is set by convention. The additional minus sign in the
definition of $v_1$ is due to the convention that $v_1(p_T)<0$ at low
$p_T$ crossing over to $v_1(p_T)>0$ at high $p_T$.

In the numerical evaluation of $dN/d^2k$ we assume $Q_s/\Lambda=10$ so
that any dependence on the infrared cutoff $\Lambda$ should be
weak. Also, we made sure that the lattice dimension $LQ_s\sim250$ far
exceeds the relevant physical scale in the problem. Lastly, in all
cases the dimensionless lattice saturation scale $Q_s^L\equiv aQ_s \ll
1$ (where $a$ denotes the lattice spacing in $r$-space) so that
discretization effects should be small. In fact, we performed a
continuum extrapolation taking $Q_s^L\to0$ at fixed $LQ_s$.

The transverse momentum distribution $dN/d^2k$ is obtained by a
Fourier transform of the dipole S-matrix as written in
eq.~(\ref{eq:Def_dN1}). We employ the semi-classical model for the
S-matrix given by
\be
\mathrm{Re}~S(\vec r) \equiv D(\vec r) =
e^{- \rho^2 \, (1-{\cal A}+2{\cal A}
\cos^2 \phi_r) \, }~, \label{eq:numerical_D_E}
\ee
and
\be \label{eq:numerical_O}
\mathrm{Im}~S(\vec r) \equiv O(\vec r) =
\frac{\alpha_s \, (N_c^2-4)}{64}
\, {\cal B} \,
\arctan \left(\frac{1}{2}rQ_c\cos\phi_r\right)
\, \rho^2\, e^{-\rho^2}
\ee
with
\be \label{eq:numerical_rho2_A=0}
\rho^2 = \frac{1}{4}\, r^2 \, Q_s^2 \, \log \left(e+\frac{1}{\Lambda
  r}\right)~.
\ee
Here, we have added a constant to the argument of the logarithm to
avoid an unphysical sign flip in the deep infrared. Below, we restrict
to dense targets $Q_s\gg\Lambda$ where the regime $r>1/\Lambda$
plays no crucial role.

We repeat that, in principle, in eq.~(\ref{eq:numerical_O})
$\phi_r\to\phi_r+\psi_0$ could be shifted by a random ``event plane
angle'', which is of no consequence for our analysis. Also, we shall
take $\alpha_s=0.25$, $N_c=3$, and consider the $\vec E$-field
polarization amplitude $\cal A$ and the strength of fluctuations $\cal
B$ as free parameters.

The main goal of this section is to show a few main qualitative
features of the equations presented above. We do not yet attempt a
quantitative description of the data. We therefore restrict to angular
harmonics for elastic quark scattering and omit the convolutions with
quark $\to$ hadron fragmentation and with the proton parton
distribution functions. Nevertheless, for a rough comparison to the
data we rescale the transverse momentum as $p_T= k_T/2$.

\subsection{Even harmonics}

\begin{figure}[htb]
\begin{center}
\includegraphics[width=10cm]{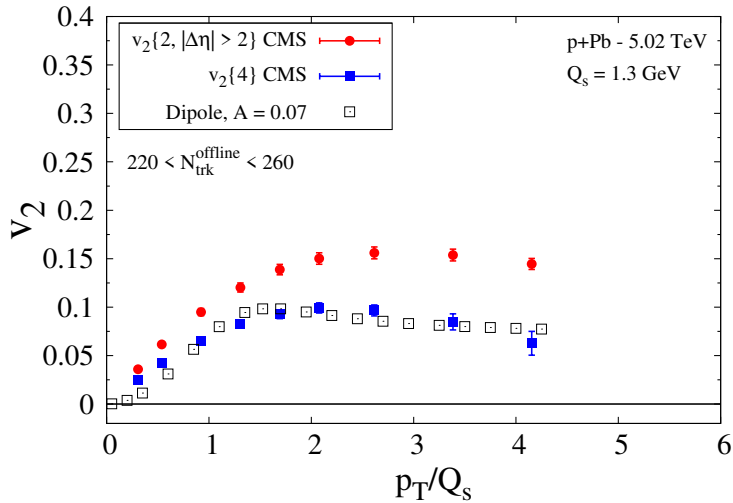}
\end{center}
\vspace*{-5mm}
\caption[a]{$v_2(p_T)$ from the semi-classical dipole model with $\vec
E$-field polarization amplitude ${\cal A}=7\%$ is shown by open
squares. Data by the CMS collaboration is shown as filled circles
($v_2\{2\}$) and filled squares ($v_2\{4\}$), respectively, and
corresponds to very high multiplicity p+Pb collisions at $\surd s=5.02$~TeV.}
\label{fig:v2_A=0.07}
\end{figure}
In fig.~\ref{fig:v2_A=0.07} we show $v_2(p_T)$ from the semi-classical
dipole model. In this plot we have scaled the experimental $p_T$ by
$Q_s=1.3$~GeV; note that this numerical value is strongly correlated
to $\langle z\rangle=1/2$ assumed in our rescaling of quark
momenta. To extract a more physical value for $Q_s$ one needs to
account for the additional convolutions with distribution and
fragmentation functions which we postpone to future work.

Despite the qualitative nature of our numerical results we may note
that with an $\vec E$-field polarization amplitude of ${\cal A}=7\%$
the model reproduces the magnitude and the rough $p_T$ dependence of
the data. Our analysis provides a first order of magnitude estimate of
${\cal A}$ which was previously unknown~\cite{Kovner:2011pe}.  Also,
with this value for ${\cal A}$ we find that $v_4(p_T)\le 1\%$ is very
small\footnote{We repeat though that here we have not performed an
  average over multiple $\vec E$-field domains explicitly, see next
  section.}, as expected from the analytical discussion in
section~\ref{sec:MVmodel}.

In fig.~\ref{fig:v2_A=0.07} we have (somewhat arbitrarily) focused on
the highest multiplicity p+Pb events analyzed by CMS. A study of the
multiplicity dependence of $v_n(p_T)$ is beyond the scope of this
work. Nevertheless, it appears that the more rapid drop of $v_2(p_T)$
(from four particle correlations) at high $p_T$ in events with lower
multiplicity may require a scale dependent polarization strength
${\cal A}$. We intend to return to this question in the future.

\subsection{Domain model}   \label{sec:Domain_Model}

In the previous section we considered a single target domain and
determined the effective polarization at ${\cal A}=7\%$. This
relatively small value of ${\cal A}$ may be due to the fact that we
are in effect averaging over multiple domains where $\vec E$ points in
different directions. This prevented us from making a prediction for
the ``quadrangular asymmetry'' $v_4$ which may average differently
than $v_2$. These questions can be addressed in a simple domain model.

One first introduces $m$-particle cumulants~\cite{Borghini:2001vi}, as
already mentioned in the introduction. For example, the two-particle
correlation is defined as
\be \label{eq:vn_cumulant_factorized}
v_n^2\{2\}\, e^{i \psi} \equiv \left< e^{in(\phi_1-\phi_2)}\right> =
\frac{1}{\cal N} \int \frac{d\phi_1}{2\pi}\frac{d\phi_2}{2\pi}
e^{in(\phi_1-\phi_2)} \frac{dN}{d^2k_1} \frac{dN}{d^2k_2} + \cdots ~,
\ee
with an obvious generalization to $m>2$ particles.
The phase $\psi$ is not important here. On the right, we have assumed
that the two-particle angular distribution factorizes into a product
of two single-particle distributions, plus corrections. Thus, if both
particles are produced from the same target domain, we obtain that
$v_n\{2\} = v_n$ as computed above. On the other hand, if (at least) one
of the particles is produced from an uncorrelated domain (with $\vec
E$ pointing in a random direction) then $v_n\{2\} = 0$.
In all, for $N_D$ domains we find that $v_n$ from $m$-particle
correlations is given by
\be
v_n\{m\} = v_n\, {N_D}^\frac{1-m}{m} ~.
\ee

\begin{figure}[htb]
\begin{center}
\includegraphics[width=8cm]{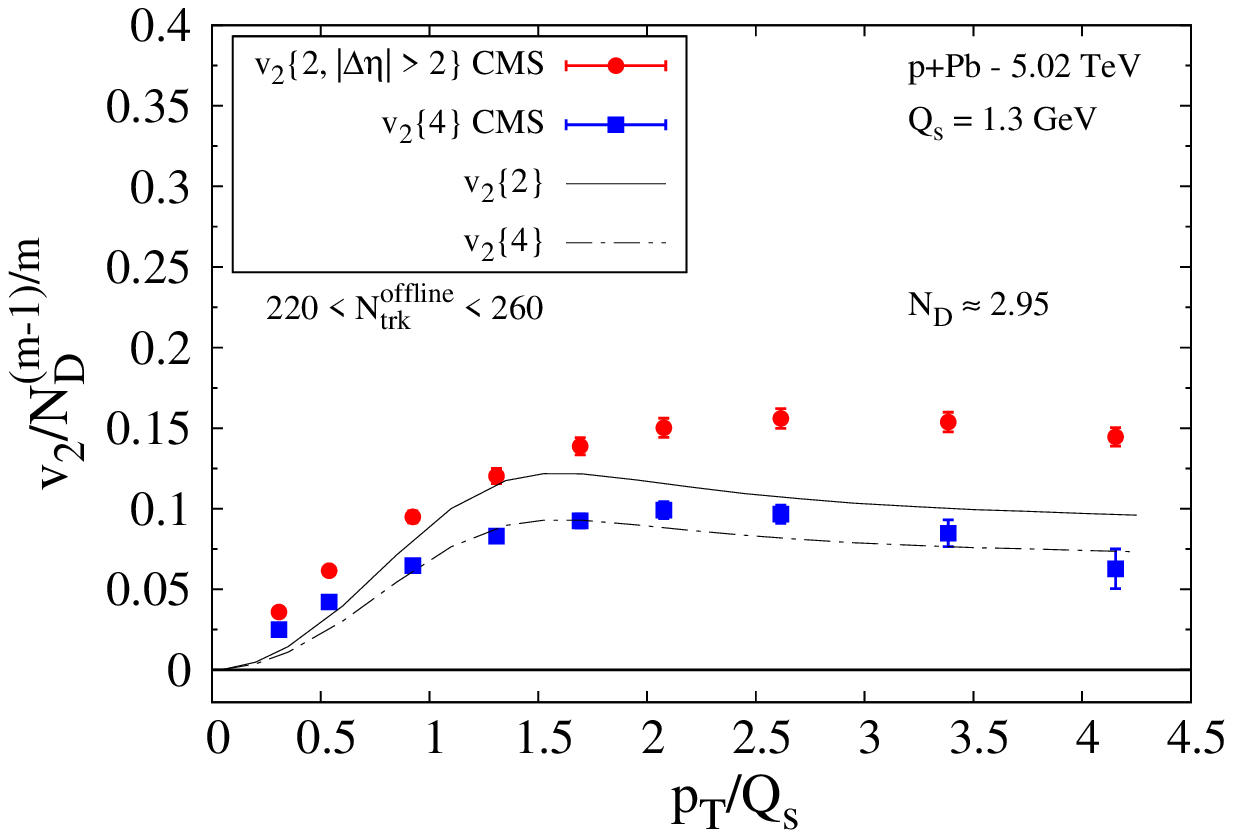}
\includegraphics[width=8cm]{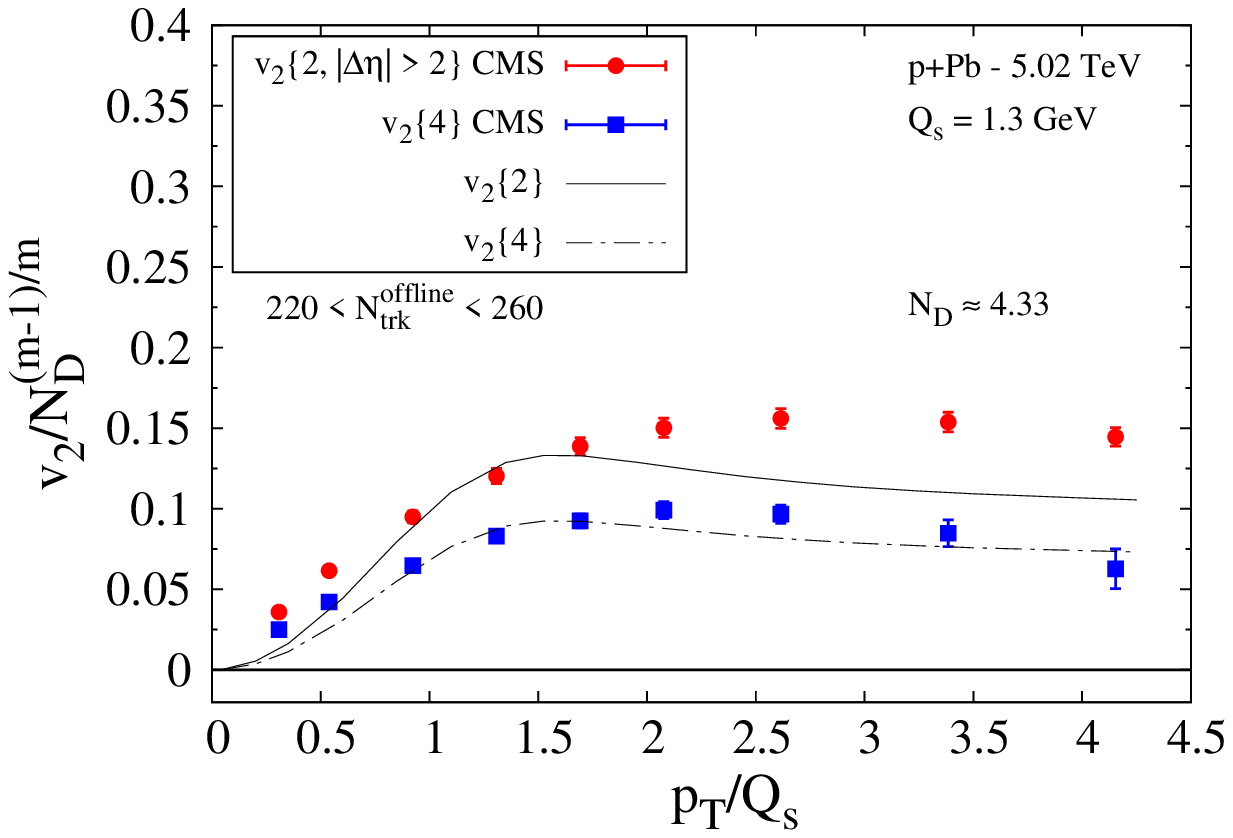}
\includegraphics[width=8cm]{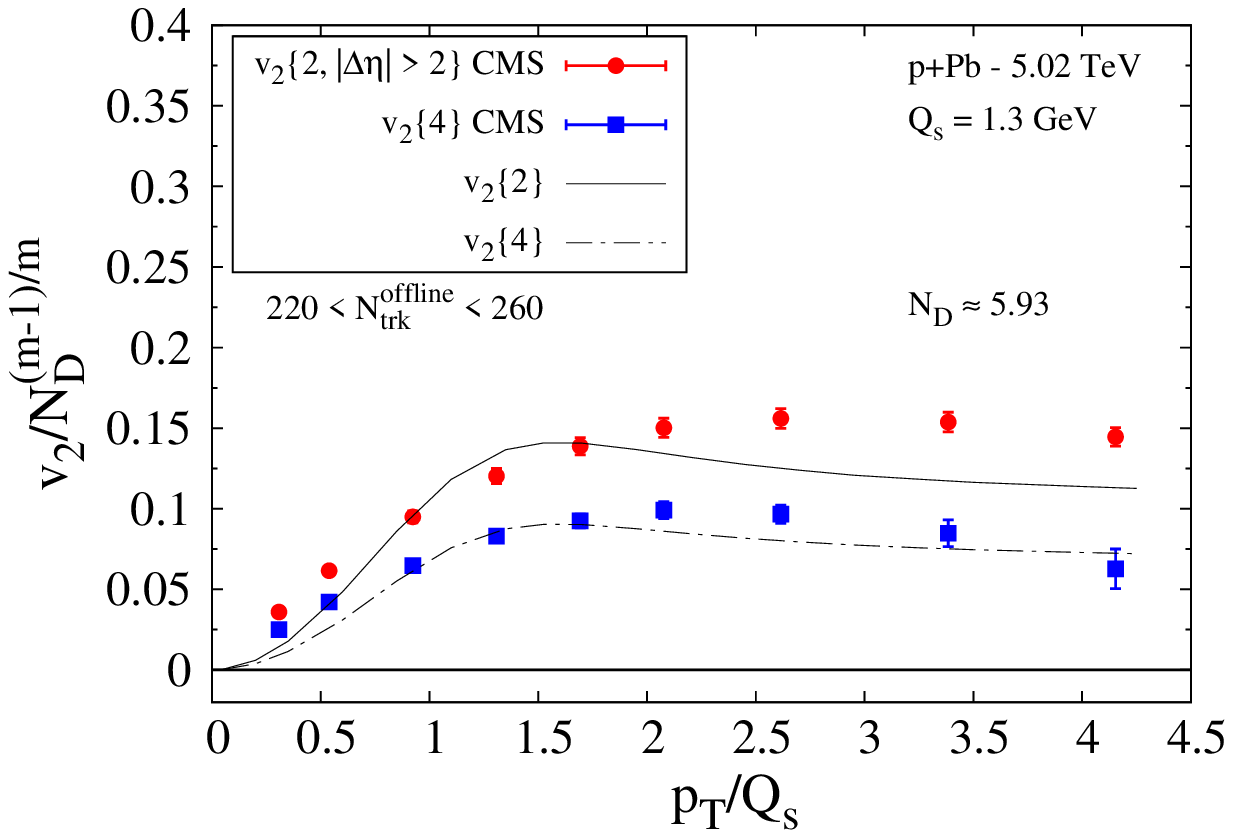}
\end{center}
\caption[a]{$v_2(p_T)$ from two- and four-particle cumulants. The
  domain model with $N_D=2.95$, 4.33, 5.93 (and, correspondingly,
${\cal A}=0.15$, 0.20, 0.25) is compared to data by the CMS collaboration.}
\label{fig:v24_domains}
\end{figure}
Fig.~\ref{fig:v24_domains} shows our fit to $v_2\{4\}$ with larger
target field polarization ${\cal A}=0.15$, 0.20, 0.25; the overall
magnitude has been adjusted via $N_D$. At the same time, we attempt to
``saturate'' as much as possible, but not to overshoot the measured
two-particle correlation $v_2\{2\}$~\footnote{We do not require a
  perfect description of $v_2\{2\}$ over the entire range of $p_T$
  since our present analysis neglects genuine two-particle
  correlations, as indicated by the dots in
  eq.~(\ref{eq:vn_cumulant_factorized}).}.

We find that the simple domain model fits best with ${\cal
  A}\simeq0.20$ and $N_D\simeq4.3$: larger values for ${\cal A}$
require larger $N_D$ which in turn results in too strong splitting of
$v_2\{2\}$ and $v_2\{4\}$ and is not compatible with the data. In this
model there appears to be some room for genuine (non-factorizable)
two-particle correlations at high $p_T$.

\begin{figure}[htb]
\begin{center}
\includegraphics[width=8cm]{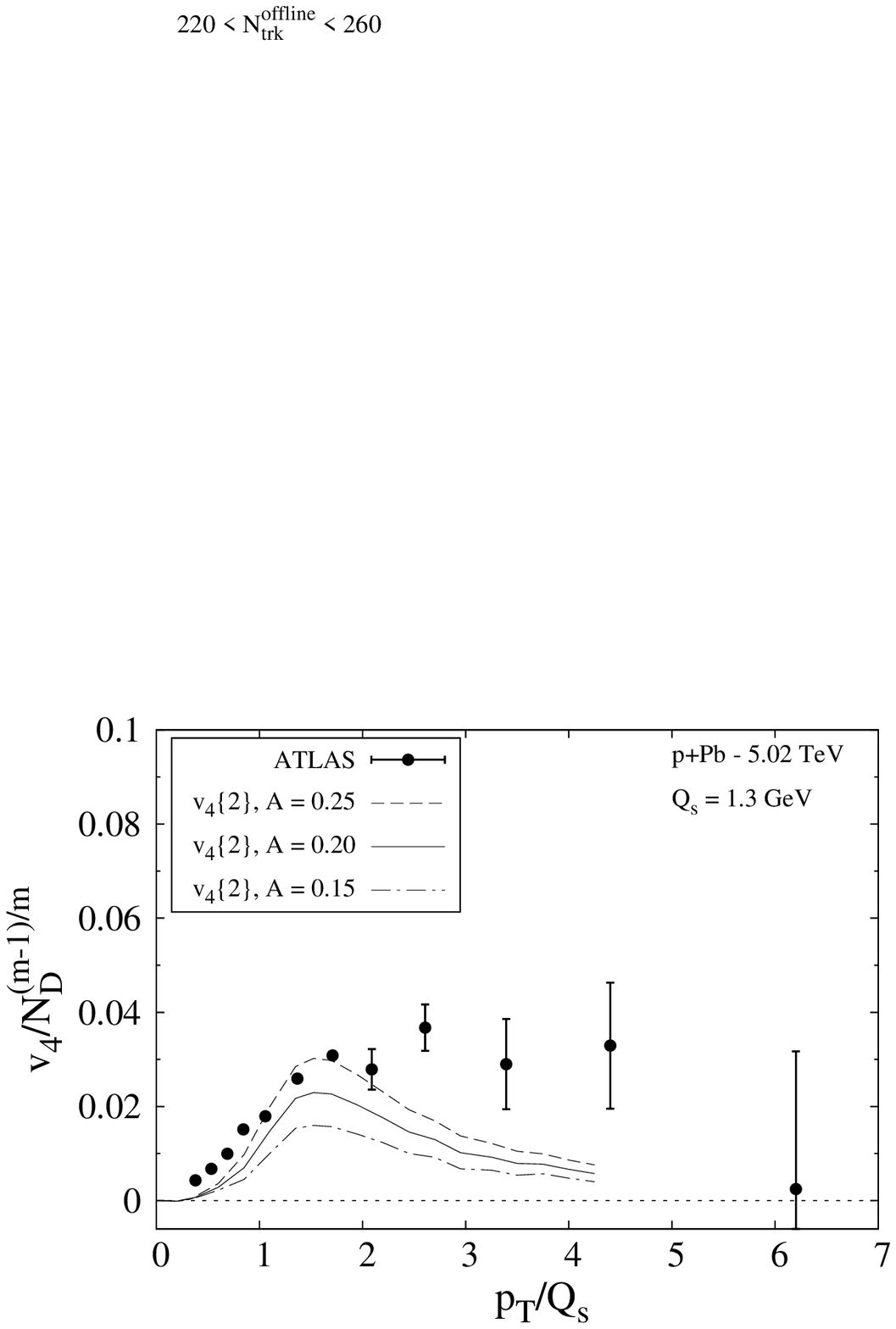}
\includegraphics[width=8cm]{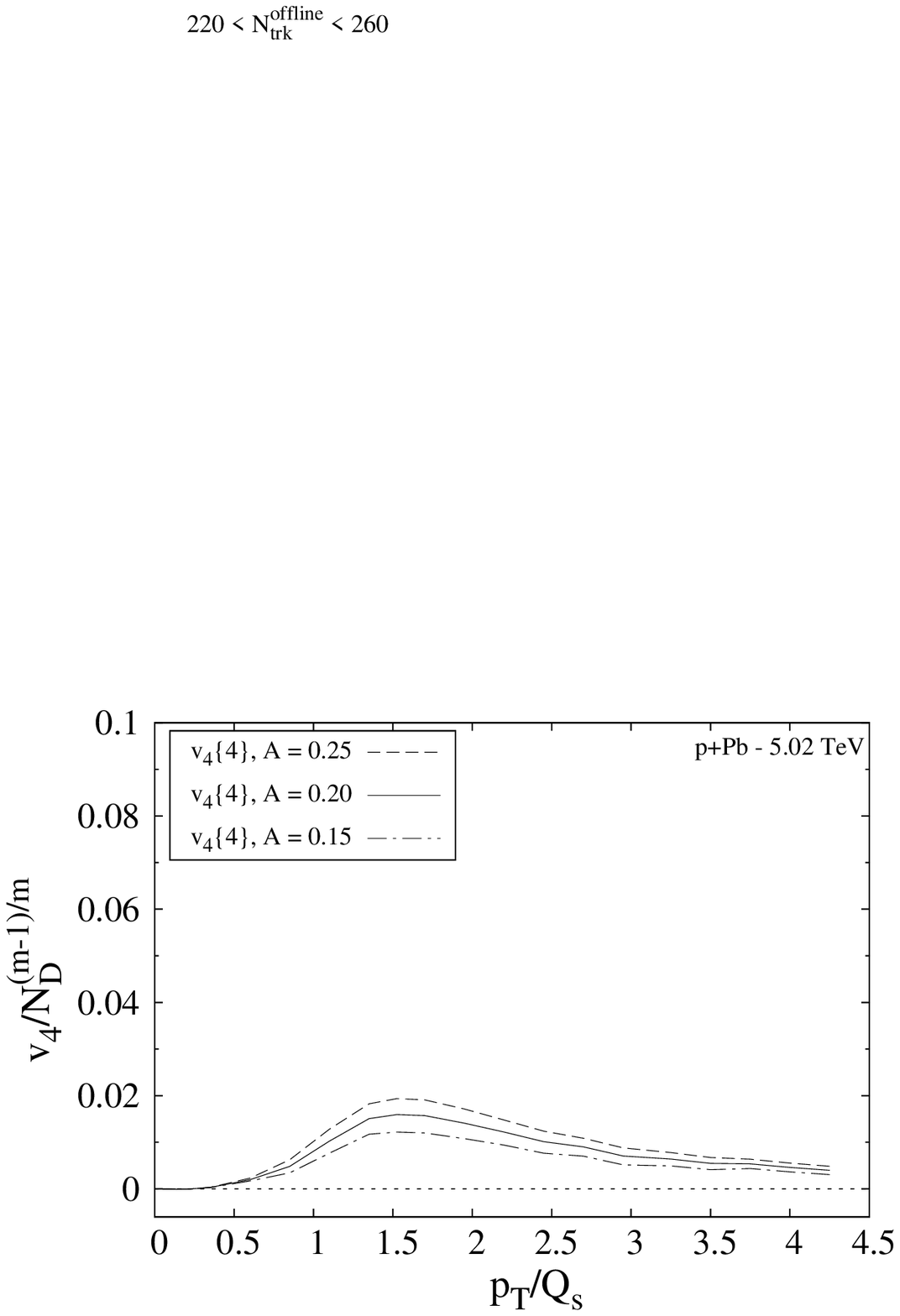}
\end{center}
\caption[a]{$v_4(p_T)$ from two- and four-particle cumulants. On the
  left, the domain model with $N_D=2.95$, 4.33, 5.93 (and,
  correspondingly, ${\cal A}=0.15$, 0.20, 0.25; from bottom to top) is
  compared to $v_4\{2\}$ data by the ATLAS collaboration. On the
  right we plot the model prediction for $v_4\{4\}$ for the same
  combinations of ${\cal A}$ and $N_D$.}
\label{fig:v4_domains}
\end{figure}
Having determined ${\cal A}$ and $N_D$ we can now use the same model
for $v_4\{2\}$ and  $v_4\{4\}$. The former has been measured by the
ATLAS collaboration and provides another test of the domain model. As
shown in fig.~\ref{fig:v4_domains}, the model does fit $v_4\{2\}$
about $Q_s$ but underestimates it at higher $p_T$. This may be
interpreted as due to the presence of non-factorizable correlations
which are not accounted for.

The prediction for $v_4\{4\}$ is shown in fig.~\ref{fig:v4_domains},
right. The model predicts a rather small $v_4\{4\} \lsim 2\%$ over the
entire range of $p_T$. More importantly, while $v_2\{4\}$ above $Q_s$
is rather flat, $v_4\{4\}$ clearly decreases with $p_T$. This is a
rather generic prediction of the $\sim \vec r\cdot\vec E$ dipole
interaction with the vector electric field.

\subsection{Odd harmonics}

\begin{figure}[htb]
\begin{center}
\includegraphics[width=8cm]{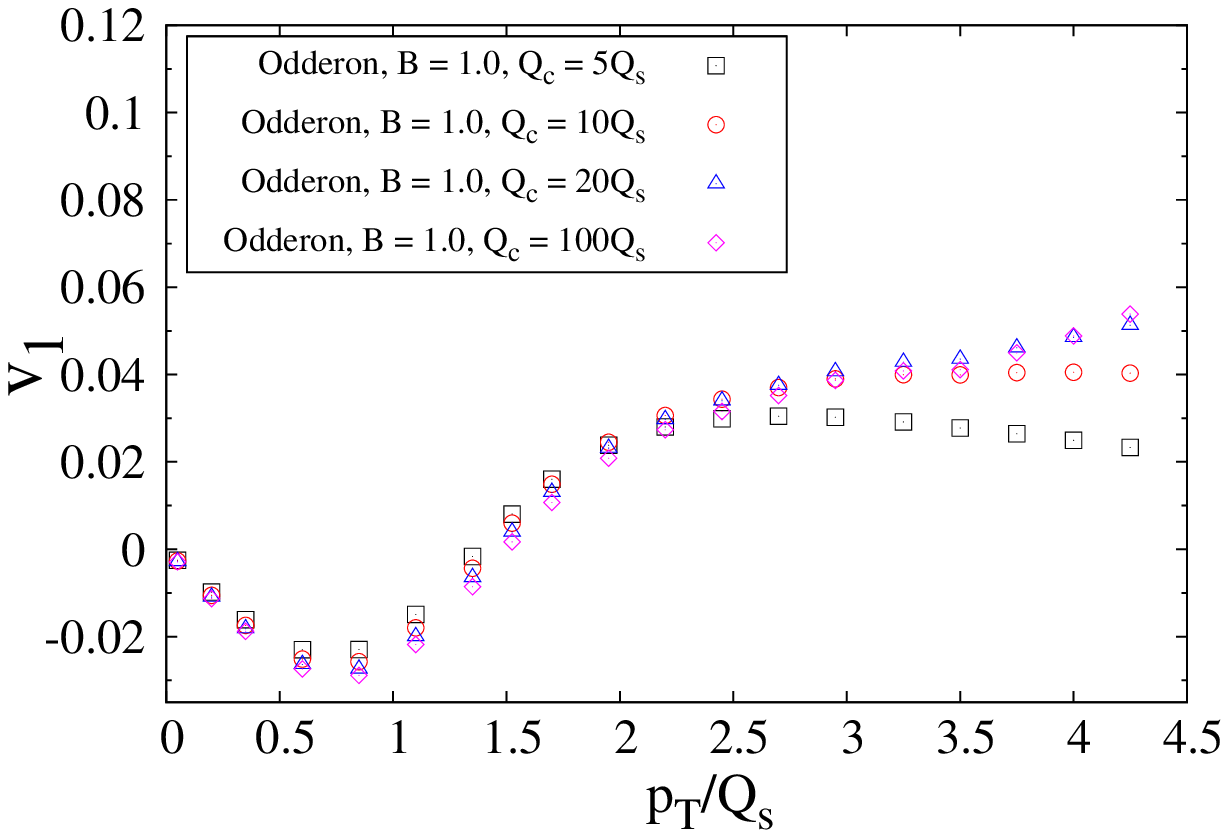}
\includegraphics[width=8cm]{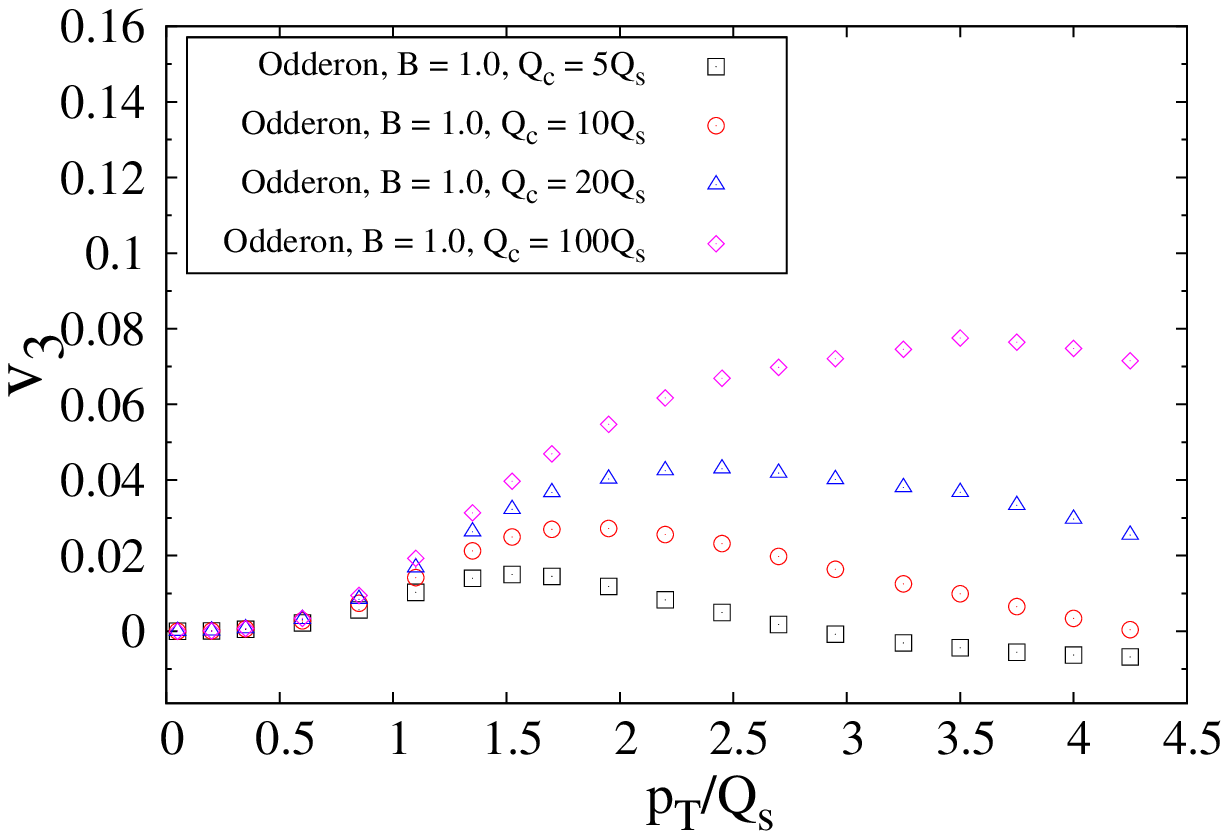}
\end{center}
\vspace*{-5mm}
\caption[a]{$v_1(p_T)$ and $v_3(p_T)$ from the semi-classical odderon
  model for a
  scale invariant distribution of C-odd fluctuations with a cutoff
  $Q_c$.}
\label{fig:v13_Qc}
\end{figure}
We now turn to odd harmonics. The $p_T$ dependence of $v_1$ and $v_3$
for different cutoffs $Q_c$ is shown in fig.~\ref{fig:v13_Qc}. For
this figure the amplitude of C-odd fluctuations has been fixed to
${\cal B}=1$ as it only affects the magnitude but not the $p_T$
dependence of the angular moments. Also, we keep ${\cal A}=7\%$ from
above and again rescale from quark to hadron momenta as $p_T=k_T/2$.

We observe that for $p_T/Q_s\lsim 2$, $v_1(p_T)$ depends rather weakly
on the cutoff $Q_c$; at high $p_T$ it increases somewhat with $Q_c$
until the cutoff far exceeds the saturation scale. On the other hand,
$v_3$ at high $p_T$ increases rather rapidly with $Q_c$. This behavior
is in line with the analytic discussion in the previous section:
fluctuations with wave length on the order of the dipole size $r$ only
produce $v_1$ but not $v_3$.

\begin{figure}[htb]
\begin{center}
\includegraphics[width=8cm]{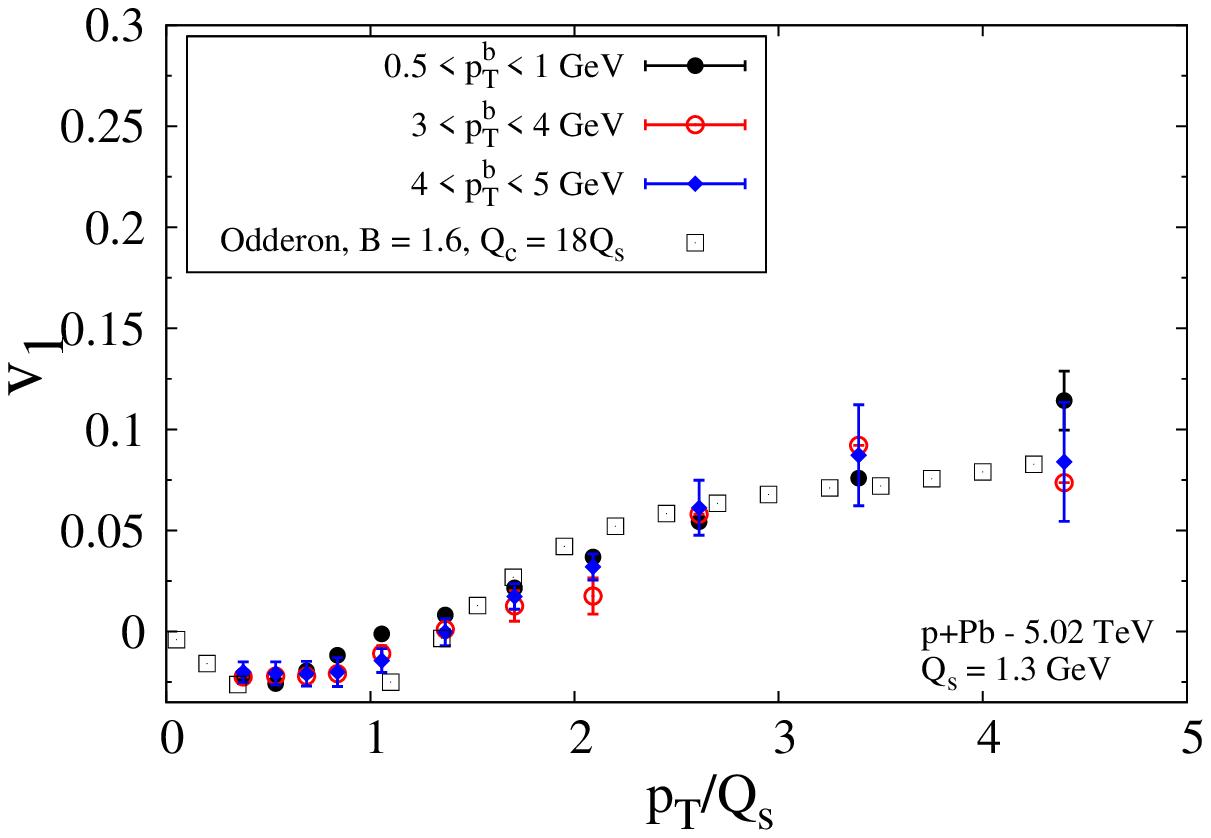}
\includegraphics[width=8cm]{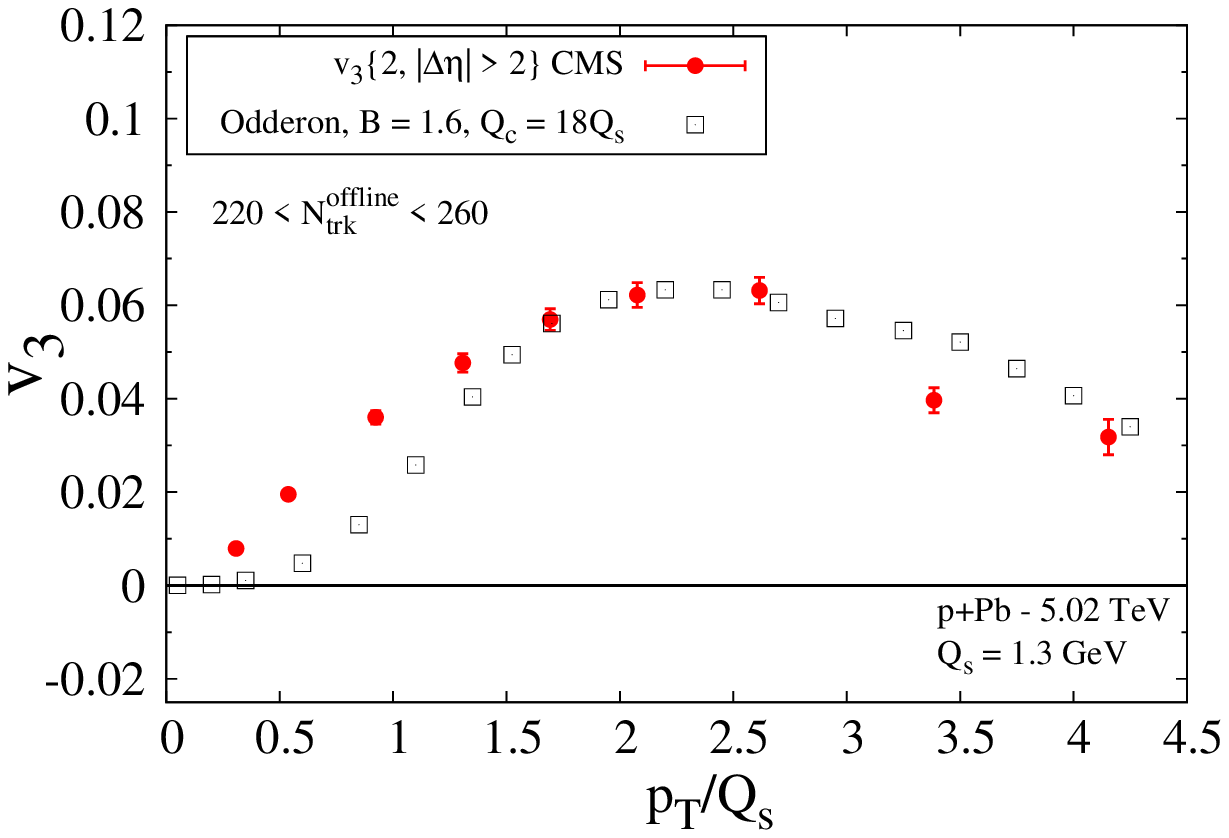}
\end{center}
\vspace*{-5mm}
\caption[a]{$v_1(p_T)$ and $v_3(p_T)$ from the semi-classical odderon
  model with fluctuation amplitude ${\cal B}=1.6$ and cutoff
  $Q_c/Q_s=18$.
Data by the ATLAS ($v_1$) and CMS ($v_3$) collaborations for
$v_{1,3}\{2\}$ corresponds to very high multiplicity
  p+Pb collisions at $\surd s=5.02$~TeV.}
\label{fig:v13_Qc=18}
\end{figure}
We can now ``tune'' the fluctuation amplitude ${\cal B}$ and the
cutoff $Q_c$ to ATLAS data for $v_1(p_T)$ from high multiplicity p+Pb
collisions at 5~TeV, and then check the behavior of $v_3(p_T)$ against
the data from CMS.  We should, of course, keep in mind that the data
corresponds to two-particle correlations $v_{1,3}\{2\}$ while the
model gives $\langle\cos n\phi_k\rangle$ evaluated on the
single-inclusive particle distribution, averaged over events.

In fig.~\ref{fig:v13_Qc=18} we show that the overall magnitude of both
$v_1$ and $v_3$ is reproduced by the odderon model with a fluctuation
amplitude of ${\cal B}=1.6$; we remind the reader that this amplitude
effectively incorporates the contribution of gluons to the denominator
in eq.~(\ref{eq:vn(k)}).

Furthermore, a cutoff $Q_c\sim 18Q_s$ leads to a decent description of
both $v_1$ and $v_3$ at transverse momenta of a few times $Q_s$. Of
course, our model provides no deeper explanation for the presence of
such hard fluctuations and it will be important to understand this
result in more detail in the future (perhaps from the Liouville
effective theory for fluctuations of $Q_s$ proposed in
ref.~\cite{Iancu:2007st}). Nevertheless, it is quite interesting that
the continued rise of $v_1$ at $p_T/Qs\gsim2$, together with the drop
of $v_3$ at such $p_T$, is basically reproduced by the odderon
model~(\ref{eq:numerical_O}) in the presence of very short wavelength
fluctuations in the target.

\section{Summary and Outlook}

In this paper we have provided a first qualitative analysis of
azimuthal asymmetries within the dipole approach. Following earlier
work by Kovner and Lublinsky~\cite{Kovner:2011pe} we have introduced a
semi-classical dipole scattering amplitude which accounts for
spontaneous breaking of 2D rotational symmetry due to a condensate for
the (color) electric field $\vec E$. Our numerical estimates indicate
that this model is able to describe the $p_T$-dependence of the
elliptic asymmetry $v_2$ obtained from four-particle correlations (or
from a global event plane method).

The second goal of this paper was to address parity odd azimuthal
moments $v_{2n+1}$. These arise from the C-odd imaginary part of the
dipole S-matrix, analogous to single transverse spin asymmetries
discussed before by Kovchegov and
Sievert~\cite{Kovchegov:2012ga}. However, in order to obtain non-zero
$v_3$ we have found it necessary to generalize the odderon to a
string-like operator~\cite{Noronha:2014vva} so that the dipole can
couple to hard fluctuations of wave length below the dipole size
$r\sim 1/p_T$. Fixing two parameters, the fluctuation amplitude ${\cal
  B}$ and cutoff $Q_c$, we find that the odderon model provides a
reasonably good simultaneous description of $v_1\{2\}(p_T)$ and
$v_3\{2\}(p_T)$ measured in high multiplicity p+Pb collisions at the
LHC. We should emphasize, however, that it would probably be more
appropriate to compare to $v_1(p_T)$ and $v_3(p_T)$ obtained from four
or more particle correlations (or from a global event plane method),
which is not currently available.

As already alluded to above, the main intention of the present work
was to explore some qualitative features of the underlying ideas. Much
work remains to be done before a more quantitative comparison to data
could be attempted. For example, we have not at all addressed the
nature of the high-multiplicity p+Pb events which display (large)
azimuthal anisotropies. Also, it will be important to employ better
models for the dipole $D(\vec r)$ and odderon $O(\vec r)$. Finally, we
will have to develop some theoretical understanding for the amplitude
and cutoff of fluctuations, which here we extracted phenomenologically.

In closing, we should point out that other possible explanations for
the observed azimuthal asymmetries in p+A collisions are presently
under intense investigation. For example, initial density
inhomogeneities of the ``fireball'' in the transverse plane could be
converted into momentum space asymmetries in the final state by
hydrodynamic flow~\cite{hydro}. Hydrodynamics is an effective theory
describing the propagation of long wavelength density
perturbations. Our present approach is complementary and relies on
short distance physics corresponding to transverse momenta far beyond
$\sim \Lambda_{\rm QCD}$. In the future, it will be important to
understand at what scale $p_T$ one transitions from one description to
the other.

\section*{Acknowledgements}
We thank K.~Dusling, V.~Skokov, and R.~Venugopalan for useful
discussions. A.D.\ gratefully acknowledges support by the
DOE Office of Nuclear Physics through Grant No.\ DE-FG02-09ER41620 and
from The City University of New York through the PSC-CUNY Research
Award Program, grant 66514-0044. A.V.G. gratefully acknowledges the
Brazilian Funding Agency FAPESP for financial support
(contract: 2013/23848-5).



\begin{thebibliography}{99}

\bibitem{pPb_ALICE}
B.~Abelev {\it et al.}  [ALICE Collaboration],
  Phys.\ Lett.\ B {\bf 719}, 29 (2013);
  arXiv:1406.2474 [nucl-ex].

\bibitem{pPb_ATLAS}
G.~Aad {\it et al.}  [ATLAS Collaboration],
  Phys.\ Rev.\ Lett.\  {\bf 110}, 182302 (2013);
  Phys.\ Lett.\ B {\bf 725}, 60 (2013).

\bibitem{pPb_v1_ATLAS}
The ATLAS collaboration,
  ATLAS-CONF-2014-021.

\bibitem{pPb_CMS}
S.~Chatrchyan {\it et al.}  [CMS Collaboration],
  Phys.\ Lett.\ B {\bf 718}, 795 (2013);
  Phys.\ Lett.\ B {\bf 724}, 213 (2013).

\bibitem{dAu_RHIC}
A.~Adare {\it et al.}  [PHENIX Collaboration],
  Phys.\ Rev.\ Lett.\  {\bf 111}, 212301 (2013).

\bibitem{Dumitru:2008wn}
A.~Dumitru, F.~Gelis, L.~McLerran and R.~Venugopalan,
Nucl.\ Phys.\ A {\bf 810}, 91 (2008).

\bibitem{Adare:2014keg}
A.~Adare {\it et al.}  [PHENIX Collaboration],
  arXiv:1404.7461 [nucl-ex].

\bibitem{Kovchegov:2012ga}
  Y.~V.~Kovchegov and M.~D.~Sievert,
  Phys.\ Rev.\ D {\bf 86}, 034028 (2012)
  [Erratum-ibid.\ D {\bf 86}, 079906 (2012)].

\bibitem{Kovner:2011pe}
  A.~Kovner and M.~Lublinsky,
  Phys.\ Rev.\ D {\bf 84}, 094011 (2011).

\bibitem{Noronha:2014vva}
  J.~Noronha and A.~Dumitru,
  Phys.\ Rev.\ D {\bf 89}, 094008 (2014).

\bibitem{Kovchegov:2002nf}
  Y.~V.~Kovchegov and K.~L.~Tuchin,
  Nucl.\ Phys.\ A {\bf 708}, 413 (2002).

\bibitem{Borghini:2001vi}
  N.~Borghini, P.~M.~Dinh and J.~-Y.~Ollitrault,
  Phys.\ Rev.\ C {\bf 64}, 054901 (2001).

\bibitem{Gyulassy:2014cfa}
M.~Gyulassy, P.~Levai, I.~Vitev and T.~Biro,
arXiv:1405.7825 [hep-ph].

\bibitem{fluc_eps_n}
A.~Bzdak, P.~Bozek and L.~McLerran,
  arXiv:1311.7325 [hep-ph];
L.~Yan and J.-Y.~Ollitrault,
  Phys.\ Rev.\ Lett.\  {\bf 112}, 082301 (2014);
A.~Bzdak and V.~Skokov,
  arXiv:1312.7349 [hep-ph].

\bibitem{Kovner:2010xk}
  A.~Kovner and M.~Lublinsky,
  Phys.\ Rev.\ D {\bf 83}, 034017 (2011).

\bibitem{pAridge}
A.~Dumitru, K.~Dusling, F.~Gelis, J.~Jalilian-Marian, T.~Lappi and
R.~Venugopalan,
  Phys.\ Lett.\ B {\bf 697}, 21 (2011);
K.~Dusling and R.~Venugopalan,
  Phys.\ Rev.\ Lett.\  {\bf 108}, 262001 (2012);
  Phys.\ Rev.\ D {\bf 87}, 054014 (2013);
  Phys.\ Rev.\ D {\bf 87}, 094034 (2013).

\bibitem{Alver:2010gr}
  B.~Alver and G.~Roland,
  Phys.\ Rev.\ C {\bf 81}, 054905 (2010)
  [Erratum-ibid.\ C {\bf 82}, 039903 (2010)].

\bibitem{Altinoluk:2014oxa}
T.~Altinoluk, N.~Armesto, G.~Beuf, M.~Martinez and C.~A.~Salgado,
arXiv:1404.2219 [hep-ph].

\bibitem{BjKS}
J.~D.~Bjorken, J.~B.~Kogut and D.~E.~Soper,
Phys.\ Rev.\ D {\bf 3}, 1382 (1971).

\bibitem{Dumitru:2002qt}
A.~Dumitru and J.~Jalilian-Marian,
Phys.\ Rev.\ Lett.\  {\bf 89}, 022301 (2002).

\bibitem{MuellerDipole}
A.~H.~Mueller,
Nucl.\ Phys.\ B {\bf 415}, 373 (1994);
Nucl.\ Phys.\ B {\bf 437}, 107 (1995);
A.~H.~Mueller and B.~Patel,
Nucl.\ Phys.\ B {\bf 425}, 471 (1994).

\bibitem{Dumitru:2005gt}
A.~Dumitru, A.~Hayashigaki and J.~Jalilian-Marian,
Nucl.\ Phys.\ A {\bf 765}, 464 (2006).

\bibitem{Altinoluk:2011qy}
T.~Altinoluk and A.~Kovner,
Phys.\ Rev.\ D {\bf 83}, 105004 (2011).

\bibitem{Chirilli:2011km}
G.~A.~Chirilli, B.-W.~Xiao and F.~Yuan,
Phys.\ Rev.\ Lett.\  {\bf 108}, 122301 (2012);
Phys.\ Rev.\ D {\bf 86}, 054005 (2012).

\bibitem{Andres:2014bia}
C.~Andr\'es, A.~Moscoso and C.~Pajares,
  arXiv:1405.3632 [hep-ph].

\bibitem{Kovchegov:1998bi}
Y.~V.~Kovchegov and A.~H.~Mueller,
  Nucl.\ Phys.\ B {\bf 529}, 451 (1998).

\bibitem{Dumitru:2013tja}
  A.~Dumitru, T.~Lappi and L.~McLerran,
  Nucl.\ Phys.\ A {\bf 922}, 140 (2014).

\bibitem{Iancu:2007st}
  E.~Iancu and L.~McLerran,
  Nucl.\ Phys.\ A {\bf 793}, 96 (2007).

\bibitem{hydro}
P.~Bozek,
  Phys.\ Rev.\ C {\bf 85}, 014911 (2012);
P.~Bozek and W.~Broniowski,
  Phys.\ Lett.\ B {\bf 718}, 1557 (2013);
  Phys.\ Rev.\ C {\bf 88}, 014903 (2013);
A.~Bzdak, B.~Schenke, P.~Tribedy and R.~Venugopalan,
  Phys.\ Rev.\ C {\bf 87}, 064906 (2013);
E.~Shuryak and I.~Zahed,
  Phys.\ Rev.\ C {\bf 88}, 044915 (2013);
P.~Bozek, W.~Broniowski and G.~Torrieri,
  Phys.\ Rev.\ Lett.\  {\bf 111}, 172303 (2013);
K.~Werner, M.~Bleicher, B.~Guiot, I.~Karpenko and T.~Pierog,
  Phys.\ Rev.\ Lett.\  {\bf 112}, 232301 (2014);
B.~Schenke and R.~Venugopalan,
  arXiv:1405.3605 [nucl-th].

\end{thebibliography}
\end{document}